\documentclass[times]{secauth}
\usepackage{setspace}
\usepackage{moreverb}
\usepackage[colorlinks,bookmarksopen,bookmarksnumbered,citecolor=red,urlcolor=red]{hyperref}
\usepackage{graphicx,amssymb,enumitem,pifont}
\usepackage{breakcites}
\usepackage{array}
\usepackage{rotating}
\usepackage{changepage}
\usepackage{amsmath} 
\usepackage{fixltx2e}
\usepackage{pifont}
\newcommand{\xmark}{\ding{53}}

\newcolumntype{C}[1]{>{\centering\let\newline\\\arraybackslash\hspace{0pt}}m{#1}}
\newcolumntype{L}[1]{>{\raggedright\let\newline\\\arraybackslash\hspace{0pt}}m{#1}} 

\newcommand\BibTeX{{\rmfamily B\kern-.05em \textsc{i\kern-.025em b}\kern-.08emT\kern-.1667em\lower.7ex\hbox{E}\kern-.125emX}}

\def\volumeyear{2013}

\begin{document}

\runningheads{M.~S.~Mohammadi}{A Privacy-Preserving Electronic Payment System for DRM}

\articletype{RESEARCH ARTICLE}

\title{A Privacy-Preserving Electronic Payment System for DRM}

\author{Mahdi Soodkhah Mohammadi\corrauth, Abbas Ghaemi Bafghi}

\begin{abstract}
One of major considerations in an online business is customer privacy. Consumers are not interested in being monitored and identified by sellers. Some solutions are proposed to hide selection of the customer but in the payment phase, there will be a leakage of information as online shopper can infer some information about customer's preference due to the price, which is paid by customer. This is a big threat to customer privacy. Our solution to this problem consists of a number of one-unit payment steps that cannot be linked to each other or to customer's identity. At the end of purchase, content provider will receive appropriate amount of money while customer will acquire a valid license anonymously. Content provider will not be able to gain any information about the customer or the content that is purchased. In addition, a dispute resolution scheme is presented for cases of conflict between customer and content provider. A series of analyses on the security, complexity and DRM requirements are presented which indicate security and practicality of our scheme.
\end{abstract}

\maketitle

\section{Introduction}
There has been a long history of efforts to protect content (a term that indicates works of art such as movie, music, books, etc.) from illegal distribution and duplication. Computers have had a strong impact on these efforts. That is because making a copy of content on a computer is almost effortless. This has been a threat to content producers and distributers. Since inception of high-speed networks this threat has become even more viable since distribution of copies of a content is now easier that before. Today with existence of different types of contents and businesses which are related to content, more advanced types of permissions (like print, lending content, valid usage period, etc.) are required.
DRM (Digital Rights Management) is a field of study in cryptography and information security, which focuses on the problem of protection of digital contents through management and enforcement of a set of permissions. Usually an online business is interested to sell a digital content with some restrictions and limitations (e.g. an online book seller may want to prevent printing an electronic book). These limitations are called permissions and a set of permissions defined for content is called a License. A customer needs a valid license that is generated by content provider in order to consume a digital content. It is common to provide customers with a special software or hardware unit (called renderer) which is required in order to consume a digital content. This unit requires appropriate license in addition to the digital content itself in order to render content for the consumer. 
In recent years there has been a strong push on development of DRM systems with specific security features and functionalities such as secure storage \cite{N1}, traitor-tracing \cite{N2, N3, N4, N5, N6, N7, N8}, watermarking \cite{N9, N10, N11, N12, N13}, fingerprinting \cite{N14, N15}, tamper resistant code \cite{N16, N17, N18} and permission revocation \cite{N6, N7, N8, N19}.

In this paper, we suppose that a customer can access content anonymously and no one is able to trace who has downloaded which content. In practice, content can be broadcasted (like IPTV \cite{N20} or satellite video) or made publicly available through a bulletin board or a public website. In case of a public website, some mechanism is required in order to make access to website anonymous, e.g., \cite{N21}, \cite{N22}, \cite{N23}.

An important focus in DRM research (which is not covered in this paper), is to propose methods to prevent illegal copy of digital content. Although there is not a working solution for this problem (because with analog output of data users can capture audio or video and sell copies of captured content) there have been some proposed methods to copy-protect digital content. These methods include legal prosecution of illegal copy distributors, watermarking techniques (which can be used to find out the source of data leak) \cite{N24, N25, N9, N10, N13, N11} and other software and hardware solutions \cite{N26, N27}.

One important issue in DRM is keeping privacy of an online customer. In fact, it is desirable for an online customer to be able to purchase a digital content without revealing any private information to any of parties, which are involved in the process. Several solutions are proposed to solve this problem. However, most of them are not practical. Some solutions force the content provider to apply a fixed price to all of contents and some others use expensive infrastructures in order to hide identity of the customer and prevent others from tracing customer's personal information. Some other solutions provide partial privacy solutions for customers in which either customer's identity or his preference is revealed.
Our contribution to solve this problem is that we have proposed a practical solution to DRM privacy. Our solution does not require a special hardware or infrastructure and the content provider is free to set any price to any content. The solution consists of a series of mini-payments size of each is one currency unit. Along with each mini-payment, a decryption key is partially constructed by content provider. At the end of payments, content provider has received appropriate amount of money while customer has a working decryption key that can be used to consume content. These mini-payments are not linkable to each other so content provider is not able to detect which content is being purchased or which step of payment is being executed at any time.

Each mini-payment is performed using a one-time prepaid card, which is bought anonymously by customer. These cards are generated by an issuing bank and can be verified at the time of purchase by receiving content provider to ensure validity of the card and prevent spending a card multiple times. 

This paper is organized as follows. Related work is covered in Section \ref{sec_related}. We present our scheme in Section \ref{sec:proposed_scheme}. In the next section \ref{sec:analysis} we provide analysis of the proposed scheme according to different approchaes: Security, Complexity and DRM requirement before we conclude and propose some open problems in Section \ref{sec:conclusion}.

\section{Related Work}
\label{sec_related}
One of important works in the area of privacy preserving payment for DRM is work of Perlman et. Al. \cite{N28}. In this paper two methods are presented to obtain a digital license anonymously. The first method is based on anonymous cash and the second one uses blind decryption. Both methods assume that there are public lists of contents, which are encrypted using content encryption keys. Potential buyers can browse content provider website to preview list of available contents and choose one of them. Each content has its own content identifier that can be used in communications with content provider to indicate the exact content that buyer is interested in.Content provider has a database of content keys and content identifiers. After purchase process, buyer will have appropriate content key, which can be used to decrypt one of publicly available encrypted contents and consume that content. 

In anonymous cash payment, the basic idea is that owning a signature of a bank on a data with a specified structure (called $R$) has a specific value that can be used for purchase operations. But if bank knows which data it has signed, this knowledge can later be used to find out identity of owner of the signature. When Alice is spending that signature as electronic cash, bank will be consulted to confirm that the signature is spent only once. Upon this consult bank can find out identity of the owner of the signature. In order to preserve identity of the signature holder, a blinding mechanism is used. In case of anonymous-cash based DRM, Buyer (Alice) provides a blinded $R$ in addition to her identity. Her identity is used for payment operation, as she needs to pay with a credit card or use her prepaid account with merchant (or bank) to pay for anonymous cash she will receive. Content provider checks received information and after successful payment, sends signature on blinded $R$ to Alice. One important drawback of this method is that we need anonymization infrastructures (which are expensive in terms of computation and bandwidth\cite{N21}). That is because buyer needs to provide content ID to the seller and without anonymization infrastructures seller can trace buyer's request and will be able to know which buyer wants to purchase the given content ID which will eliminate privacy of buyer. Another drawback of this method is that buyer needs to reveal her identity to content provider upon receiving anonymous cash. That is because content provider needs to know which account should be debited (in case of using pre-paid accounts with seller) or, in case of credit card payments, seller needs to know billing information of buyer.

In Blind Decryption based payment DRM, Alice sends blinded encrypted key to the content provider who will decrypt this data. As data is blinded, content provider will not be able to understand any information. Blind decryption is similar to blind signature but as it does not require a "public key" there are more algorithms for blind decryption than blind signature. Blind decryption can be used with various schemes including RSA, Diffie-Hellman and IBE (identity based encryption).This method has two major drawbacks. First issue is that seller is able to understand identity of buyer. That's because buyer needs to provide a signature on her request. Another issue is that, in most cases seller will be able to indirectly deduce some information about the content, which is purchased. Because buyer needs to provide the exact amount of money she wants to pay for a content. If seller knows amount of money he is receiving, in case of contents with varying prices, he will understand the content that is being purchased. The only case that we can prevent this problem is by forcing seller to set a fixed price for all contents, which is not applicable in real world scenarios. A solution for this problem is proposed (which we call Enhanced blind decryption based DRM) in the paper  which suggests using multiple keys each with multiple values. 

Conrado et. Al. \cite{N30} present a privacy-preserving DRM scheme which is based on smart cards. Major entities that are involved in this scheme are user, smart card (SC), smart card issuer (SCI), compliance certificate issuer for smart card (CA-SC), compliance certificate issuer for compliant device (CA-CoD), compliant device (CoD) and content provider (CP).

The paper explains three protocols for content purchase, anonymous license transfer and authorized domain creation.
In content purchase scheme, a user first buys a smart card from a retailer that provides anonymous smart cards. Each smart card has an asymmetric key-pair (PK, SK) and a preset PIN number. User does not learn value of SK as the only entity who knows this value is SC. Upon content purchase, a user contacts a content provider (CP) through anonymous channel and provides a receipt of anonymous payment for content price in addition to PK of smart card. CP will be able to link a PK to content but this does not threat user's privacy as his identity is not linkable to PK. CP encrypts license with PK and signs the result and sends the response to SC. As data is signed by PK, only SC will be able to decrypt this data.

In order to participate in interactions with a compliant device, SC needs a compliancy certificate from a CA-SC. To acquire this certificate, sends PK to a compliancy certificate issuers who checks validity of the PK. There is a revocation list of PKs available from SCI, which includes list of public keys, which are revoked (due to illegal activity or transfer of license). After this check, CA-SC creates a random pseudonym called RAN and sends signed $H(RAN)$ in addition to encrypted RAN under PK to SC. This certificate can be later used as a proof of compliancy of smart card without revealing personal information or PK stored in smart card. One disadvantage of using a compliancy certificate in multiple cases is its likability. In order to resolve this problem, DRM rules enforce adding a validity period to a compliancy certificate and periodical renewal of this certificate.

In order to consume a digital content on a compliant device (CoD) first user needs to transfer encrypted content and license to the device. It is assumed that a CoD contains required compliancy certificate, which can be provided to user's smart card to prove compliance of the device. Smart card presents its compliancy certificate to CoD to prove its compliance too. When a content is being consume, device can capture identity of the user but the only link that device is able to create is between user identity and smart card compliance certificate which is not useful as compliance certificates are renewed frequently.
CoD sends encrypted license to SC for decryption using SC's private key. After decryption, CoD can extract content decryption key from license and play the content for user. 
In case of attacker's access to CoD it is possible to link a user's identity to $RAN$ (used in SC's certificate), content and license. However, attacker will not be able to learn $PK$. 

The second protocol enables a user to transfer a license between to another user anonymously. In order to perform a license transfer, first user (whose SC's public key is PK) sends a request to content provider including his public key, license details and public key of receiving user (PK'). Content provider, revokes first user's license and creates a new license for the second user. In order to prevent linkability of public keys it is proposed to use a generic anonymous license. Upon transfer request, public key of receiving user is not sent to content provider. Instead an anonymous license is generated which is sent to requesting user. This license is sent to second user who contacts content provider to convert this license to a personal license. A method based on blind signature is used in order to prevent content provider being able to link a previously generated anonymous license with public key of second user.

Usually a DRM user shares a set of devices with other members of a group (e.g. a family). In order to handle this scenario, a concept of authorized domain is introduced. Authorized domain is a collection of CoD's which are shared between members of a group each having his own licenses and contents. A method is explained to create and manage a domain, which preserves privacy of group members and prevents other entities to learn structure of the group.

This main drawback of the schemes proposed in this paper is that it forces users to possess a trusted hardware device (which comply with certain DRM rules) in order to be able to participate in the scheme. In addition, it is assumed that users contact authorities through anonymous channels, which seems impractical in World Wide Web. 

In \cite{ANEWT} Kavitha et. al. have proposed an transferable electronic cash which preservers privacy of the digital cash holder. This scheme is based on Proxy Re-signature scheme which is used to support transfer of the money for a limited number of times. The system uses a trusted third party to prevent double-spending of electronic money. Spending of digital cash is done through a transfer protocol which is run between two users. At any point, a holder of digital cash can provide his coins to the bank and deposit associated amount of money into his account. 

Juans \cite{ANEFFFAIR} has proposed an efficient and fair exchange scheme which preserved anonymity of the buyer. The scheme is based on bilinear pairings. An exchange scheme is an electronic commerce scheme which provides exchange between payment of the customer and the digital goods of the merchant. One important feature of this system is a dispute resolving phase which is executed after exchange. This phase is divided into two sections. The first section deals with customer requesting help and the second is for merchant requesting help. In this scheme, bilinear pairings and elliptic curves are used to reduce computation and communication costs. 

Wang et. al. \cite{ANIMOFF} have proposed an off-line electronic cash scheme which preserves privacy of spenders. In this paper work of Ziba et al. \cite{Eslami201159} is analyzed and some security flaws are found which prevent revealing identity of double-spenders. The proposed scheme fixes flaws of Ziba's system. Furthermore, this scheme supports all desirable properties of electronic cash and is based on blind signature and discrete logarithm. Since this scheme is based on multiple authorities (Bank and Central Authority) each of which contains a set of private information which are related to the customer, in case of cooperation between parties, privacy of the scheme will be at risk.

Win et al. \cite{PRIVENDRM} have proposed a privacy enabled DRM without reliance on a trusted third party. This scheme uses blind signature and one-way hash functions to eliminate need for a trusted third party. The system uses a hierarchical network of content providers which act as a seller of digital goods to end-users. The entity that creates a digital good is called the owner which uses content provider network to distribute and sell digital content. This scheme uses anonymous tokens to preserve privacy of the customers. These tokens are generated by content owner and end users can acquire them anonymously to be used during content purchase. 

Zhang et al. \cite{LMSAT} have proposed a license mangement scheme for DRM which uses anonymous trust (LMSAT) to preserve privacy of the license buyer. This scheme uses elliptic curve cryptography and is divided into two phases: License acquisition phase and usage tracking phase. The paper assumes that each end-user has a DRM Agent (DA) installed on his personal computer which acts on behalf of the user to acquire a license and consume the digital content. Other important participants of LMSAT are Content Producer, Content Provider, Clearing House and the Client. During license acquisition phase, a DRM agent acquires a valid license which is bound to anonymity ID from Clearing House. In this phase it is assumed that the user has paid for the content and an anonymity ID is acquired as a result of payment phase.

\subsection{Comparison}
Table \ref{tbl_comparison} represents a comparison between our scheme and two categories of privacy based DRM methods in terms of factors, which are important in privacy. In this table, $\circ$ means that the scheme meets the requirement incompletely or partially.

\begin{table}
\caption{{\small Comparison of related works}}
\begin{tabular}{|L{1.1cm}|C{.35cm}|C{.35cm}|C{.35cm}|C{.35cm}|C{.35cm}|C{.35cm}|C{.35cm}|}
\hline 
\begin{sideways} \textbf{Title/Feature} \end{sideways} & 
\begin{sideways} \textbf{Buyer Identity Privacy} \end{sideways} & 
\begin{sideways} \textbf{Purchased Content Privacy} \end{sideways} &  
\begin{sideways} \textbf{Support for variable prices} \end{sideways} &  
\begin{sideways} \textbf{Does not require ANETs} \end{sideways} &  
\begin{sideways} \textbf{Privacy Preserving in case of cooperation} \end{sideways} &
\begin{sideways} \textbf{No reliance on TTP} \end{sideways} &
\begin{sideways} \textbf{Dispute Resolution} \end{sideways} \\  

\hline \cite{N28} {\footnotesize (AC)} & \checkmark  & \xmark & \checkmark & \xmark & \checkmark & \checkmark & \xmark \\ 
\hline \cite{N28} {\footnotesize (BD)} & \xmark & \checkmark & $\circ$ & \checkmark & \checkmark & \checkmark & \xmark \\ 
\hline \cite{N30} & $\circ$ & \checkmark & \checkmark & \xmark & \xmark & \checkmark & \xmark \\ 
\hline \cite{ANEWT} & \checkmark & \xmark & \checkmark & \checkmark & \checkmark & \xmark & \xmark \\ 
\hline \cite{ANEFFFAIR}  & \checkmark & \xmark & \checkmark & \checkmark & \xmark & \xmark & \checkmark \\ 
\hline \cite{ANIMOFF}  & \checkmark & \xmark & \checkmark & \checkmark & \xmark & \checkmark & \xmark \\ 
\hline \cite{PRIVENDRM}  & \checkmark & \xmark & \checkmark & \checkmark & \checkmark & \checkmark & \checkmark \\ 
\hline \cite{LMSAT} & \checkmark & \xmark & \checkmark & \checkmark & \checkmark & \checkmark & \checkmark \\ 
\hline
\end{tabular} 
\label{tbl_comparison}
\end{table}

\section{Proposed Scheme}
\label{sec:proposed_scheme}
\begin{figure}
\centering
\includegraphics[width=\linewidth]{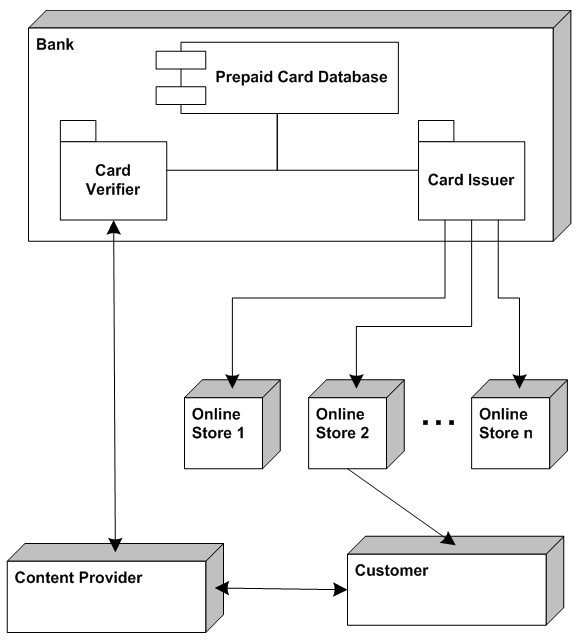}
\caption{Proposed scheme: static view using UML deployment diagram}
\label{fig_static}
\end{figure}

In order to provide payment infrastructure for our scheme, we use a system, which is called prepaid card system. Figure \ref{fig_static} illustrates static structure of players in the prepaid card system and the path of data flow between those entities. Bank generates, stores and distributes prepaid cards to online stores. A customer will buy prepaid cards from online stores to be able to participate in purchase process with a content provider. 

\begin{figure}
\centering
\includegraphics[width=\linewidth]{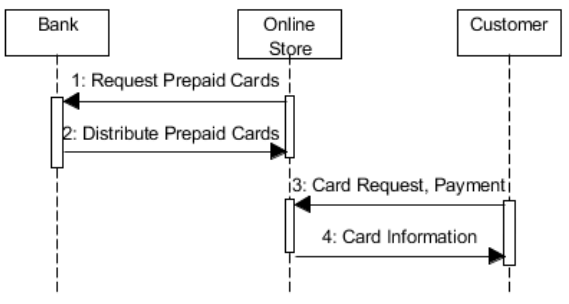}
\caption{Proposed scheme: dynamic view using UML sequence diagram (Card Distribution)}
\label{fig_dynamic_card}
\end{figure}

Our scheme has two main operating phases. First phase is called card distribution and the second phase is purchase. In card distribution phase, bank distributes cards' information to online stores and in exchange receives associated amount of money. No content purchase is performed in this phase. Figure \ref{fig_dynamic_card} represents actions that are taken place in this phase.

The second phase (purchase phase) is the focus of this paper. In this phase a purchase operation is performed. To perform a purchase operation, a customer sends information of one of his prepaid cards to a content provider in addition to a request. Content providers contacts bank to check validity of  the given card and if validation succeeds, it sends appropriate response to the customer. The verification step is done in order to prevent a user from spending a single prepaid card multiple times. As all cards are generated by a single entity (issuing bank) and all spent cards are checked with that same entity, in case of double spending, the bank will detect the fraud and inform the accepting shop about this. Please note that only one card is spent in this step so the purchase operation is not finished after verification is successful.

\begin{figure}
\centering
\includegraphics[width=\linewidth]{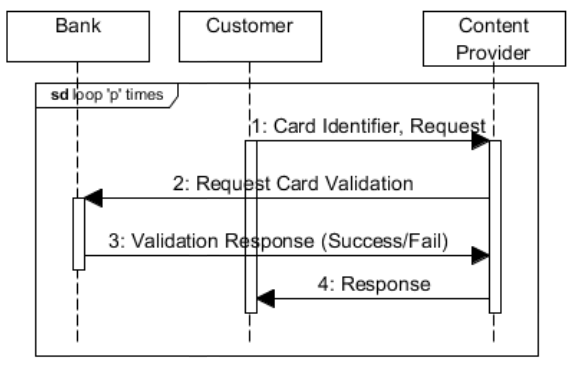}
\caption{Proposed scheme: Dynamic view using UML sequence diagram (Purchase)}
\label{fig_dynamic}
\end{figure}

This process is repeated '$p$' times where '$p$' is the price of the digital content that customer wants to purchase. In this mode, content provider acts as a stateless entity so it does not detect end of a series of requests from a customer regarding a single purchase and is not able to link multiple requests to each other. The details of this step are explained in "Purchase Scheme" section. Figure \ref{fig_dynamic} represent overview of this mode.

One useful property of using prepaid cards as anonymous electronic cash is that identity of the customer is not revealed during purchase of a prepaid card or spending a prepaid card. As a result, privacy of a customer is preserved while at the same time payment operation between customer and an online shop (content provider) can be done securely and confidently.

\subsection{Purchase Overview}
Purchase operation of our scheme can be considered as an enhanced version of anonymous-cash DRM that does not require Anonymizing Networks and can keep privacy of purchased content. This scheme can be called "Blinded Partial Decryption" in the sense that the decryptor (content provider) does not know the information that is being decrypted and performs a partial decryption in each step. After a series of 'p' partial decryptions ('p' being price of the content), original data will be available to the customer.

\subsubsection{Prerequisites}
We assume that an online seller (content provider) has a number of digital contents available for sale. Contents can have any number of licenses attached. Each license has its own license identifier. In order to consume content, one needs to obtain a valid license for that content. This process is called license purchase in which buyer provides seller with appropriate amount of digital money and seller provides buyer with a valid license decryption key that can be used to decrypt an encrypted license.
In addition, we assume a public website that contains information about contents and licenses. Encrypted contents are public and can be anonymously downloaded. In order to consume content, one needs to provide one of licenses for that content. Encrypted licenses are publicly available to download.

Digital contents can have different licenses for different types of consumers. For example for a digital book, we can have two licenses. One which only allows reading the book and another license which enables consume to print the book in addition to reading it. The major difference between licenses is their price, which reflects differences in the features provided by them.

In addition, we suppose existence of a one-time prepaid card system in which a consumer can buy any number of prepaid cards anonymously by paying appropriate amount of cash.
These cards do not reflect identity of their owner and they can be used only once in a payment process. The most basic method to implement this feature is by banks generating unique identifier strings and stores them in a Card Database. Each string in the Card Database can be used to print a prepaid card for public use. Each card has a value of one currency unit (e.g. one dollar or one cent) although it is possible to publish multiple-value prepaid cards. These cards can be distributed to physical shops and consumers can buy them and use their identifier strings in purchase operations.

Each license has a set of license terms that are made public. Seller signs these terms plus encrypted license. This signature is public which will be used in case the encrypted license is not according to publicly stated terms. In case of dispute between seller and buyer, this signature can be used to prove a third party about invalid claims of the seller about license terms. The complete process is explained in the "Dispute Resolution" section.

We assume existence of a secure channel for information exchange between customer and content provider.

\subsubsection{Scheme Overview}
In the purchase operations, buyer provides seller with identifier string of his receipt and seller can check this identifier with bank in order to confirm its validity.
After selecting content and one of its licenses, buyer needs to prepare appropriate number of prepaid cards for payment. 

For purchase operation, buyer blinds license identifier and provides seller with blinded license identifier in addition to identifier string of his first prepaid card. 
Seller verifies validity of given prepaid card string identifier. If validation fails, seller sends appropriate response indicating error message and purchase process is terminated. In case of successful validation, bank credits seller's account. Seller performs an exponentiation on blinded license identifier and returns the result to the buyer.
After doing above process '$p$' times (where '$p$' is license price), seller will have '$p$' currency units in his bank account and buyer will have a decryption key which can be used to decrypt license and consume content.

\subsection{Scheme Details}
\label{sec:scheme_details}
Proposed scheme consists of two phases; Setup phase and Purchase phase. In the setup phase, some global system parameters are generated and system data structures are published for customers' view. In the purchase phase, we have explained process of license acquisition and payment operations.

\subsubsection{Setup Phase}
First phase of scheme is setup phase. In this phase, seller prepares some important public and private parameters of the scheme. For brevity we use '$B$' to denote buyer and '$P$' to denote content provider (seller) in this scheme.

This phase contains below steps:

\begin{enumerate}
\item $P$ selects a large prime number '$n$'. All calculations in this scheme are modulus '$n$'.
\item $P$ creates a private/public key-pair for a secure public key encryption scheme. The private key will be used for signing operation and public key will be used to check $P$'s signature.
\item '$s$' is $P$'s private license generation factor which is randomly selected.
\item Each content ($Content_j$) has a number of licenses: $L_1,L_2,...,L_n$.
\item Each license ($L_i$) has a public license encryption factor which is denoted by $x_i$. No two $x_i$ numbers should be powers of each other, e.g. $\forall i, \nexists j \neq i;x_i=x_j^a$.
\item Each license $L_i$, is encrypted using $C_i=x_i^{s^{p_i}}$, where $p_i$ is license price. This is indicated by $E_{C_i}(L_i )$ where $E$ is a symmetric encryption algorithm (e.g. AES or any secure encryption algorithm). 
\item For each license, a signature of $P$ is made public. This signature is performed on the encrypted license and license terms. This will be used in the dispute resolution scheme in the case that decrypted license (at the end of scheme execution) does not confirm with publicly stated license terms. 
\item $P$ selects a public generator $g$.
\item Value of $K_1=g^s$ is made public.
\end{enumerate}
	
\begin{figure}
\centering
\includegraphics[width=\linewidth]{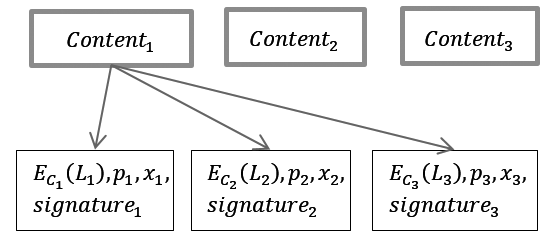}
\caption{Overview public data structures for a seller with three available contents, first content has three licenses each with its own price}
\label{fig_datastructure}
\end{figure}

Figure \ref{fig_datastructure} represents a sample seller's public data structures.

\subsubsection{Purchase Phase}
When $B$ wants to purchase a license $L_i$, with price $p_i$, he needs to buy '$p_i$' prepaid cards beforehand. Purchase operation is performed according to following steps:
\begin{enumerate}
\item $B$ selects a private random $\alpha$ and calculates $r=g^\alpha$. r is blinding factor which will be used to prevent $P$ from understanding which content, $B$ wants to purchase.
\item $B$ sends first prepaid card identifier in addition to $M_1=rx_i$ to $P$.
\item $P$ confirms validity of prepaid card (using issuer bank's online services) and then raises received $M_1$ to s and returns $M_1^s$  in addition to a signature on $[M_1,M_1^s]$. This signature will be used in a dispute resolution process in case of conflict between buyer and seller.
\item $B$ checks received signature. In case of a corrupt signature, a dispute resolution process (type C) will be executed. Details of this process are explained in "dispute resolution" section. $B$ divides received response to $K_1^\alpha$ in order to calculate $x_i^s$.
\item $B$ sends another prepaid card identifier and $M_2=rx_i^s$.
\item $P$ confirms validity of prepaid card and returns $M_2^s$ in addition to a signature on $[M_2,M_2^s]$. This signature will be used in a dispute resolution process in case of conflict between buyer and seller.
\item $B$ checks received signature. In case of a corrupt signature, a dispute resolution process (type C) will be executed. Details of this process are explained in "dispute resolution" section. B divides received response to $K_1^\alpha$ in order to calculate $x_i^{s^2}$.
\item After '$p_i-2$' rounds of execution of steps 5 through 7, $B$ will be able to calculate value of $C_i$.
\item $B$ can use $C_i$ to decrypt related license and consume the content.
\end{enumerate}
	
\begin{figure}
\centering
\includegraphics[width=\linewidth]{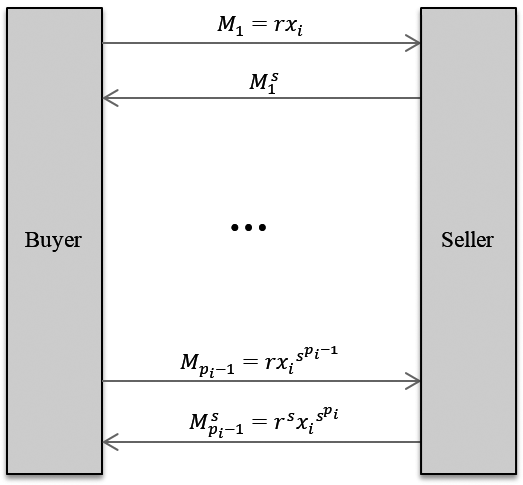}
\caption{Overview of messages in our scheme}
\label{fig_messages}
\end{figure}

Figure \ref{fig_messages} represents overview of message transmission in our scheme (excluding prepaid card identifiers).
At the end of protocol execution, we have following conclusions:
\begin{enumerate}
\item Seller has received $p_i$ valid prepaid cards where $p_i$ is value of the license which is purchased.
\item Buyer has received $C_i$ which can be used to decrypt appropriate license and use the related content.
\item Seller does not have any information about identity of the buyer.
\item Seller does not have any information about the license which is sold because any information which is received by the seller is blinded (multiplied by $r$) or anonymous (using prepaid cards).
\end{enumerate}
	
First two items ensure correctness of the scheme and the second two items ensure that scheme keeps privacy of the buyer against seller.

To buy a license with price $p_i$, $2p_i$ message transmission between seller and buyer is required. This does not seem a high number but for a famous seller with a large number of customers this can impose a high bandwidth and computation overload which can impact his online reputation.

In order to fix this problem, we can modify the scheme as follows:
In the setup phase, $P$ publishes values for $K_i=g^{s^i}, i=1,...$. During scheme execution, value of $K_2$ can be used instead of $K_1$ to merge two payment messages into one message. In this case, B will send a normal blinded data in addition to value of two prepaid card identifiers (or a prepaid card which has a value of two currency units). 
When $P$ receives a message, after verification of prepaid cards, he will raise received message $M_i$ by $s^2$ hence the response will be $M_i^(s^2 )$. $B$ can divide received response by $K_2^\alpha$. This procedure can happen for any other value $i$ for which there exists $K_i$.

$P$ can publish any number of $K_i$s to decrease required bandwidth and number of messages that need to be transmitted for a purchase operation. Our suggestion is to publish values for $K_1$, $K_2$, $K_4$, ..., $K_{2^n}$ where $n=\lfloor log_2(\max{p_i})\rfloor$. In this case, buying a license with price $p_i$ will require at most $\lceil log_2{p_i} \rceil$ message transmissions.

\subsection{Scheme Features}
\label{sec:useful_features}
A useful feature of our scheme is zero-cost license upgrade. Suppose that customer has acquired a content license which costs '$p$' currency units. Related content has another license which costs '$q$' where $q>p$ (e.g. a basic license which costs a small amount of money and a more advanced license which is more expensive). If user wants to upgrade his license to the advanced license, all he needs is to perform purchase operation on his license decryption key for '$q-p$' steps. After these steps, customer will have an upgraded advanced license. The only requirement is that content provider should have used a common license encryption factor $(x_i)$ for both types of licenses.
Our scheme supports variable priced digital goods and licenses. Many other schemes force seller to set fixed price for all of his contents to ensure privacy of the buyer. This limitation is not practical in real-world scenarios and decreases chances of applicability of those schemes.

Usually a digital store has a number of digital contents for sale. Nevertheless, each content  can have different licenses for different types of usages of that good. This is a useful feature that is not covered in any related work. Supporting this feature makes a scheme more practical and more interesting for online sellers. That is because a digital good can have different features attached to it. For example a set of permissions are available for a digital book. These permissions include ability to lend the book, print the book pages, copy the book, etc. In a real world online bookstore, seller wants to offer different prices for a book, which has printing feature enabled, or a book, which can be lent by the buyer to others. As a result, seller is interested to sell a single book with different prices each with its own set of permissions. In DRM, we call a set of permissions related to a digital content, License. Therefore, we need to support a variable number of licenses for a single digital book. Our scheme covers this feature and lets a seller define any number of licenses per digital good.

\subsection{Dispute Resolution}
\label{sec:dispute}
One major problem in an electronic commerce related scheme is how to deal with disputes. A number of probable disputes can occur which if not resolved appropriately using a fair method can make a good scheme impractical. A dispute is any kind of disagreement between two parties of the scheme (buyer and seller) which cannot be resolved appropriately. To resolve a dispute, two parties need to contact a trusted third party (an arbitrator or a judge) and provide related evidence. The arbitrator will then process given evidence and claims of two parties to declare result. 

Four possible disputes can happen in our scheme, which are called type A, B, C and D disputes.
Type A dispute happens when a seller states that a buyer has acquired a license without paying for it. This type of dispute cannot happen according to security features of the scheme. A seller will continue execution of the scheme only in the case of receiving valid prepaid cards as payments. Without proper execution of the protocol, a buyer will not be able to extract any information about the license and decryption keys. The explanation of the details of this feature can be found in the "security analysis" section.
Type B dispute happens when a buyer states that after successful execution of the purchase operation and receiving a valid and working license-decryption key, the actual terms and permissions of the license are not according to seller's claims.

Resolution of a type B dispute is straightforward, as seller has committed to his claims before scheme starts. This commitment is publicly available using signature of the seller on license terms and encrypted license. A buyer can provide arbitrator with seller's signature, encrypted license and a log of protocol steps in order to be verified and processed. If seller's signature is valid, protocol steps are correctly performed and the result, decrypted license, does not agree with committed terms, arbitrator states that seller is responsible for a refund. In any other case, buyer's claim will not be accepted. Please note that in this process, buyer will need to provide some private information to the arbitrator in order to prove his claim. Although doing this will help arbitrator check the claims the buyer has made but buyer's privacy will be damaged.
Third type of dispute happens when buyer claims that he has received a corrupt or invalid signature from seller in the middle of scheme execution. In this case buyer will contact an arbitrator and provide request/response in addition to the received signature. Arbitrator checks signature's validity and if buyer's claim is verified, asks seller to provide originally signed values. If values provided by seller and buyer are the same, arbitrator asks seller to generate a valid signature for provided original values. In case of conflict between these values, this dispute becomes a type D dispute in which seller has to prove arbitrator that his original data are correctly constructed. If seller's proof is accepted by arbitrator, buyer's claim is rejected and seller's provided data will be sent to the buyer. Please note that here we have an unfair advantage for seller, as he is able to convince arbitrator using self-generated numbers, but we assume that a seller is interested in increasing his reputation by making customers satisfied. Providing a solution for this problem will decrease anonymity of the customer, which is not desirable. Because of mentioned reasons, we ignore this problem. If seller fails to convince arbitrator that he has acted honestly, arbitrator asks seller to provide a signature on values provided by the buyer.
The fourth type of dispute (type D) occurs when buyer claims that after honest execution of scheme steps, at the end he has not received a working key and is not able to decrypt the appropriate license.

To resolve this case, buyer needs to contact arbitrator for each step of the scheme. Suppose that transmitted messages in the kth step of the scheme are, $Q_k=rx_i^{s^{k-1} },N_k=(rx_i^{s^{k-1}})^s$ where $Q_k$ is the request message sent from buyer to seller and $N_k$ is appropriate response. For each step a dispute resolution process needs to be done. For each execution of this process, buyer provides arbitrator with $Q_k$,$N_k$ and seller's signature on these values. This signature is used by arbitrator to confirm that seller approves sending given response as a result of that request. If checking of any of steps is failed, arbitrator will declare that seller has to provide correct value to the buyer. In case that all steps are validated without any issue, arbitrator declares that seller is honest and buyer's claims are rejected.
Three methods can be used to resolve this type of dispute. The first method which is the safest one in the sense that no private information of the seller needs to be sent to arbitrator, consists of a zero-knowledge proof protocol execution \cite{N36, N37}. In this protocol, seller acts as prover and arbitrator acts as a validator. Prover needs to prove that:

\begin{center}
$log_{N_k}(Q_k)=log_g{K_1}$
\end{center}

Values of $K_i$ are public. These values are used in the purchase process for unblinding data but they are commitment of seller on value of s. So this feature can be used to prove honesty of the seller to the arbitrator.

The second method uses a license decryption key, which is a private information of the seller. In this method, arbitrator selects a random license, $L_i$,from available licenses and asks seller to provide its decryption key, $C_i$, using a secure channel. After checking validity of this key, arbitrator asks seller to perform a zero knowledge proof \cite{N37, N38}  that:

\begin{center}
$log_{N_k}(Q_k)^{p_i}=log_{x_i}(C_i)$
\end{center}

Here $p_i$ and $x_i$ are price and license generation factor of the selected license, respectively. His zero knowledge proof, shows that seller has used the same value for exponentiation of buyer's request which is used to create license decryption key $C_i$. This method provides a better proof for seller but some private information need to be sent to arbitrator.

The third method requires seller to send his private value, $s$, to the arbitrator. Arbitrator will re-generate request response messages to check whether seller has performed honestly or no. This method uses seller's private data, which is a negative point, but the result will be deterministically correct.
In any type of dispute resolution, neither seller nor arbitrator will have access to buyer's private information. Seller does not receive any type of information. The only responsibility of a seller in dispute resolution is providing proof of his honesty using zero knowledge. Also arbitrator will not gain any information about the identity of the buyer or the license which is being purchased because arbitrator does not know which step is being processed right now and also all information which is sent to arbitrator (request and response) are blinded, which cannot be used to extract any information about license which is being purchased.

\section{Analysis}
\label{sec:analysis}
In this section we analyse our scheme according to different approchaes. At first we evaluate requirements of a DRM system and the way our scheme addresses them. Then we analyse complexity of our scheme (explained in section \ref{sec:scheme_details} and figure \ref{fig_messages}) in terms of computations and communication overhead for scheme participants. After that a discussion is presented about security and anonimity of our scheme. 
\subsection{DRM Requirements Analysis}
\label{sec:drm_req}
A DRM system has a number of stakeholders such as the content provider, rights provider and the end user. Each stakeholder has its own incentives to use a DRM system. If incentives of participants are not met correctly, this system will not be useful and applicable. Below we discuss our scheme with regards to DRM fundamental security requirements as mentioned in \cite{N39}. 
\begin{enumerate}
\item Content should be available only in encrypted format using a secure encryption algorithm and decryption keys are separately distributed by Rights Provider.
\item An interoperable, well-defined and fine-grained rights expression language is indispensable. In addition, license should be transferrable to other devices within authorized domain of devices under control of the customer.
\item Content and license should be securely distributed and transmitted.
\item Information transmission between Content provider and Rights provider should be over a secure channel.
\item Content provider should use some specific mechanisms (e.g. Watermarking \cite{N9}) in order to embed some hidden data structures within content for copyrights protection and pirate prosecution.
\item A DRM system should protect privacy of the user.
Here we present a discussion about requirements that are covered in our scheme.
\end{enumerate}

\subsubsection{Content Encryption}
We assume that contents are available in encrypted format. In order to consume content, a customer needs three elements: Encrypted content, License and renderer software.
An encrypted content is the content that customer wants to purchase. In order to prevent illegal distribution of content, it is provided in encrypted format. Content will be decrypted on customer's machine upon consumption.

A license includes content decryption keys in addition to license terms. 

A renderer software is a tool, which enables a customer to make use of a content. For example, this can be a movie player, sound player or electronic book reader software.

\subsubsection{Secure Content Transfer}
As we have supposed that content provider and Rights Provider are same entities, our scheme does not require secure data transmission between these two entities.
Another aspect of this requirement is secure distribution and transmission of content and license between content provider and customer. Our scheme meets this requirement as all public data (content and license) are in encrypted format and we have assumed a secure channel between customer and content provider. This secure channel can easily be created using a secure key agreement protocol (e.g. Diffie-Hellman key exchange \cite{N40}) or by customer sending a predefined symmetric key to content provider secured using content provider's public key.

\subsubsection{Customer Privacy}
Our scheme preserves privacy of the customer in the sense that no personal information of the customer is sent to any party of the system. In order to purchase a content, customer does not need to disclose his identity, billing information or even preference. No authority (content provider, online shops or bank) will be able to conclude what content is bought by customer, even if they collude together.

\subsection{Complexity Analysis}
\label{sec:complexity}
\subsubsection{Computation Complexity}
According to seller's point of view, the computation workload is quite low. Seller always receives messages in format (A,B) where A is information about a prepaid card or a set of prepaid cards and B is a blinded number. Seller needs to validate 'A' information and if confirmed, raise 'B' to his private license generation factor, $s$, in order to prepare response message. 

In the basic scheme, on the buyer side, a random number, $\alpha$, is selected and two one-time calculations are done. First $R=g^\alpha$ is done to calculate blinding factor, then $K_1^\alpha$ is calculated which will be used to unblind future messages. Then for each step of the scheme, one unblinding operation (division to $K_1^\alpha$) is done.
Table \ref{table:basic_communication}, represents computation complexity of the basic version of the proposed scheme for a license purchase with value p.

\begin{table}
\caption{Computation Complexity of the Basic Scheme}
\begin{tabular}{|c|c|c|}
\hline \textbf{Entity/Calculations} & \textbf{Buyer} & \textbf{Seller} \\ 
\hline Exponentiation & $2$ & $p$ \\ 
\hline Division & $p$ & $0$ \\ 
\hline Signing & $0$ & $p$ \\ 
\hline Total & $p+2$ & $2p$ \\ 
\hline 
\end{tabular} 
\label{table:basic_communication}
\end{table}

In enhanced scheme, multiple exponentiations are made on the seller's side and buyer needs to calculate multiple values of $K_i^\alpha$ for different values of i. Table \ref{table:enhanced_communication} represents computation complexity in the worst case for enhanced version of the proposed scheme for a license purchase with value p. 

\begin{table}
\caption{Computation Complexity of the Enhanced Scheme}
\begin{tabular}{|c|c|c|}
\hline \textbf{Entity/Calculations} & \textbf{Buyer} & \textbf{Seller} \\ 
\hline Exponentiation & $1+log(p)$ & $p$ \\ 
\hline Division & $log(p)$ & $0$ \\ 
\hline Signing & $0$ & $log(p)$ \\ 
\hline Total & $1+2log(p)$ & $p+log(p)$ \\ 
\hline 
\end{tabular} 
\label{table:enhanced_communication}
\end{table}

\subsubsection{Communication Complexity}
One of widely used mechanism to provide privacy protection in a payment system is using ANETs. Although these systems hide identity of participants, they impose a lot of overhead in terms of bandwidth and computations. Our scheme preserves privacy of customers without using ANETs.
\par Each request message from buyer to seller in the basic scheme contains information of a prepaid card and a blinded number. Response will be a number too. If we suppose that prepaid-card identifier string has $\beta$ bits and scheme modulus has $\gamma$ bits, for a purchase of a license worth p, the communication overhead of the protocol for buyer will be $p(\beta+\gamma)$ bits and for seller this overhead will be $p\gamma$ bits plus size of digital signatures which is supposed to be $\gamma$ for each signature.
In the enhanced scheme, the above calculations will be much smaller as in the worst case we will be using $log(p)$ instead of $p$. 
Table \ref{table:commproposed}, represents communication complexity of our scheme:

\begin{table}
\caption{Communication Complexity of the Proposed Scheme}
\begin{tabular}{|c|c|c|}
\hline \textbf{Entity/Scheme version} & \textbf{Buyer} & \textbf{Seller} \\ 
\hline Basic version & $p(\beta + \gamma)$ & $2p\gamma$ \\ 
\hline Enhanced version & $log(p)(\beta + \gamma)$ & $2log(p)\gamma$ \\ 
\hline 
\end{tabular} 
\label{table:commproposed}
\end{table}

\subsection{Security Analysis}
\label{sec:security_analysis}
In order to do a security analysis on the privacy-preserving feature of our scheme, let suppose that seller has found out identity of content buyer. This means that the seller is able to trace buyer's identity using his prepaid card identifier. This conflicts with our assumption that process of purchase of prepaid cards is done anonymously.
Now suppose that seller has found out the license (hence the content) that buyer is interested into. This means that given $rx_i$ seller is able to extract value of $x_i$. As a result seller has the knowledge about value of r which conflicts with our assumption in the first step of purchase order.
Suppose that buyer has calculated value of seller's private license generation factor. In this case, he will be able to calculate any license without paying money and without any type of contact with seller. This means that either he knows value of s in advance (which conflicts with assumption that s is private) or he is able to calculate value of s during scheme execution. The latter case means that given $x_i^s$ and by having value of $x_i$, he is able to calculate value of $s$. In order to solve this equation, he needs to solve discrete log problem for which no polynomial algorithm exists \cite{N41, N42}.

As a result we can conclude that our scheme is secure in the sense that seller is not able to infer any information about buyer and buyer cannot calculate private system parameters.

The only data that seller will receive is prepaid card data and a blinded string. None of these data can be used to infer any information about identity of the buyer, the license in which buyer is interested or how many payments are done so far for that license. Seller is not able to understand if received message is the first message of a series of payments for a license or last message or any of intermediate messages. As a result, we have achieved a higher level of blinding which we call "Blinded Security Protocol". In a blinded security protocol, one party does not know the current step of protocol execution. According to all of his information, he is not able to deduce whether protocol is just started or is finishing. This feature enhances privacy of the other party (buyer in our scheme).

One important parameter of the scheme is license generation factor, $s$, which is used by content provider to encrypt licenses. If this number is stolen from content provider, security of the whole system will be at risk. To solve this issue, content provider can regenerate a new license generation factor and content licenses and also expire old licenses. In this case, no user will be able to consume an expired license. Of course, legitimate users who have previously purchased these licenses can use them as long as the original validation period is not passed.

In the purchase process steps, customer selects a blinding factor (R) which is used to blind request sent to content provider. In order to increase security and privacy of the customer, it is possible to change value of R in each step. This enhancement will impose a computational overhead to customer but at the same time, security against side channel attacks (e.g. IP routing monitoring and message eavesdropping) will be increased.

\section{Conclusion and Further Work}
\label{sec:conclusion}
We have come up with a privacy preserving electronic payment system, which can act as a component of a DRM system. This payment system helps buyers purchase digital goods without revealing their identity or preference to the seller or financial institutions. Our scheme provides complete anonymity of the buyer while at the same time, makes seller sure that he receives full payment of digital goods. Our scheme does not impose any limitations on the pricing of digital goods and also does not require an anonymization infrastructure (ANET). A Dispute resolution scheme is presented which uses a trusted third party to resolve conflicts between buyer and seller without threatening privacy of the buyer.
\par One important problem in DRM systems is tracking traitor users. A traitor is a user who acts against DRM system rules (e.g. distributes copies of a digital good to others). In order to track such users, a DRM system needs mechanisms to identify a traitor and revoke his license. 
One important step of our scheme is the prepaid card verification step, which requires online connection of seller with card issues that is generally a bank. This requirement may impose a high overhead to the seller and the bank. We believe that enhancing our scheme with an offline verification system can have a significant impact on it's applicability.

\bibliographystyle{wileyj}
\bibliography{References}

\end{document}